\documentclass[preprint,12pt,authoryear]{elsarticle}

\usepackage{amssymb}
\usepackage{multirow}
\usepackage{float}
\usepackage[colorlinks=true, allcolors=blue]{hyperref}
\usepackage{booktabs}
\usepackage{tabularx}
\usepackage[utf8]{inputenc}
\DeclareUnicodeCharacter{2212}{-}
\usepackage{amsmath}

\journal{Nuclear Physics B}

\begin{document}
	
	\begin{frontmatter}
		
		
		
		\title{A Bayesian Statistical Study of Bianchi Type-I Universe in $f(R,T^\psi)$ Modified Gravity} 
		
		
		\author[label1]{Mohit Thakre} 
		
		\affiliation[label1]{organization={Symbiosis Institute of Technology, Nagpur Campus, Symbiosis International (Deemed University)},
			addressline={}, 
			city={Pune},
			postcode={440008}, 
			state={Maharashtra},
			country={India}}

        \author[label1]{Praveen Kumar Dhankar} 

            \author[label2]{Safiqul Islam} 
		
		\affiliation[label2]{organization={Department of Basic Sciences, General Administration of Preparatory Year, King Faisal University},
			addressline={P.O. Box 400, Al Ahsa 31982, Saudi Arabia \& Department of Mathematics and Statistics, College of Science, King Faisal University}, 
			city={P.O. Box 400},
			postcode={Al Ahsa 31982}, 
			state={},
			country={Saudi Arabia}}
            \author[label3]{Parbati Sahoo} 
		
		\affiliation[label3]{organization={Department of Mathematics, Bhadrak Autonomous College},
			addressline={FM University}, 
			city={},
			postcode={756100}, 
			state={Odisha},
			country={India}}
\author[label4]{Farook Rahaman} 
		
		\affiliation[label4]{organization={Department of Mathematics, Jadavpur University},
			addressline={}, 
			city={Kolkata},
			postcode={700032}, 
			state={West Bengal},
			country={India}}
\author[label5, label6]{Behnam Pourhassan} 
		
		\affiliation[label5]{organization={School of Physics, Damghan University},
			addressline={}, 
			city={},
			postcode={3671641167}, 
			state={Damghan},
			country={Iran}}

		
		\affiliation[label6]{organization={Center for Theoretical Physics, Khazar University},
			addressline={41 Mehseti Street, Baku}, 
			city={},
			postcode={AZ1096}, 
			state={},
			country={Azerbaijan}}
		\begin{abstract}
			We have examined the cosmological actions of LRS ( Locally Rationally Symmetric ) Bianchi type-I universe model in $f(R,T^\psi)$ gravity. For this, we have estimate Hubble parameter, effective equation of state parameter ($\omega^{\text{eff}}$) and potential of scalar field as a function of time using equation $H=W(\psi)$. The graphical representation of potential function $V(\psi)$ with respect to cosmic time $t$ is described. This study explores the dynamical properties of a Bianchi Type-I universe by utilizing Bayesian statistical techniques to constrain the model parameters and evaluate the viability of anisotropic cosmology under extended matter-geometry couplings. Also, we have applied Markov Chain Monte Carlo (MCMC) mechanism on derived $H(z)$ model by using  observational Hubble data (OHD), Baryon Acoustic Oscillation (BAO) dataset and Pantheon dataset. From the confidence–level contours and best–fit parameter values obtained, along with the corresponding reduced $\chi^{2}$, it is evident that the model aligns strongly with observational data, demonstrating statistical stability and consistency in describing late–time cosmic acceleration. Likewise, the error analyses presented in this research, including a comparison between the $\Lambda$CDM cosmology and the reconstructed $H(z)$ model, confirm the model's compatibility with current observations by yielding a reliable and accurate account of the universe's expansion history.
		\end{abstract}

		
	\end{frontmatter}
	
		
		
		\section{Introduction}
		\label{sec1}
		
		A significant turning point in contemporary cosmology is the discovery of the universe’s accelerated expansion, which profoundly shapes modern understanding of its fundamental structure. This accelerated phase is generally attributed to the mysterious component known as dark energy, whose origin and properties remain elusive. The phenomenon is well described by the $\Lambda$CDM model, where the cosmological constant $\Lambda$ acts as the source of dark energy within the framework of General Relativity (GR). However, the $\Lambda$CDM model faces two major theoretical challenges: the coincidence problem, which questions why the vacuum energy density is of the same order of magnitude as the matter density, and the fine-tuning problem, arising from the enormous discrepancy between the observed value of $\Lambda$ and its theoretical prediction from quantum field theory. These issues have motivated the development and study of modified theories of gravity as viable alternatives to GR, offering promising explanations for cosmic acceleration.

The forthright of general relativity, the $f(R)$ theory surrogates its general functions f(R) for the Ricci scalar $R$ in the Einstein Hilbert action. A more generic theory such as $f(R,T)$ gravity was proposed by Harko \textit{et al.}~ \cite{1harko2011f}, where $T$ is the trace of the energy momentum tensor. Gravity in which Jamil \textit{et al.}
\cite{2jamil2012reconstruction} suggested few cosmic models and found that about reproduces the $\Lambda $CDM model. For precise model of gravity, Houndjo~\cite{3houndjo2012reconstruction} talked about matter concurred and accelerated period of the universe. After studying thermodynamics Sharif and Zubair~\cite{4sharif2012cosmol} came to inference that second law of thermodynamics relates to both phantom and non phantom phases.

The $f(R, T^{\psi})$ was also put by Harko \textit{et al.} in the work \cite{1harko2011f}. One probable source of dark energy which is a solar field, which has a function same as a gas under negative pressure. It could be traits of a force field that accountable for the universe's expansion \cite{5longair1996book}. Halliwell give detailed thought about this in \cite{6halliwell1987scalar}. In their inquiry of the effect of scalar fields of gravitational lensing, Virbhadra \textit{et al.}~\cite{7virbhadra1998role} found some new findings. In addition with arguing a number of cosmologically major cases, Bazeia \textit{et al.}~\cite{8bazeia2006first,9bazeia2015first} introduced first order formalism for solving differential equations for scalar field models. On the huge basis, our universe is homogeneous and isotropic and flat in space, still new experimental documentation suggest that the universe is anisotropic and leans to become isotropic over time ~\cite{10bennett2003microwave,11de2004significance,12schwarz2004low}. The classes which are isotropic in nature of a Riemannian manifold or Bianchi classification were posed by Bianchi ~\cite{13bianchi1894lezioni}. In the structure of general relativity (GR), primary research on Bianchi models was put forward in ~\cite{14taub1951empty,15witten1962gravitation}. In the context of general relativity and modified theories to explore the anisotropic foundation of cosmos, Bianchi type models have been broadly studied ~\cite{16singh2008bianchi,17sharif2009exact,18wilson2010loop,19reddy2013some}. The Bianchi type I model which is spatially flat, homogeneous anisotropic version of the FRW spacetime that is the most honest. Aftur studying this model in Brans-Dicke theory, Sharif and Waheed~\cite{20sharif2012anisotropic} came to the inferencethat the anisotropic fluid leans to become isotropic after on which is coherent with the most convetional observational documentation. Bianchi type I model which is the locally rotationally symmetric (LRS) of warm affectation was interrogated by  
~\cite{21sharif2014warm} who showed that the model is cooperative with observational data.A full cosmological picture in $f(R, T^{\psi})$ gravity for a homogeneous and isotropic universes was lately examined by ~\cite{22moraes2016complete}. In this study. we construct on this work by using LRS Bianchi type I model to examine how anisotropy impacts physiological parameters. The layout of this paper is accordingly. In the following section, we use first order mechanism to express the field equations and find the values of $H$, $\omega_{\mathrm{eff}}$, and $V(\psi)$.

An turning point in concurrent cosmology was the disclosure of universes accelerated expansion at the twist of the twenty first century. The first mighty suggestion that the expansion rate of the universe is increasing rather than decreasing , as was earlier thought given the domination of matter and radiation came from observations of far Typela Supernovae(SNe la) \cite{29riess1998observational,38schmidt1998high,39perlmutter1998discovery}. These creative studies operated separately by the High-$z$ Supernova search team and all Supernova cosmology project, established the presence of dark energy(DE), an exotic type of negative pressure energy that records for almost 70\% of background (CMB) radiation and BAO observations have further supported this phenomena and when combined, they described a consistent picture of an accelerating universe \cite{32spergel2003first,33komatsu2009five,101ade2016planck,88beutler20116df,89anderson2014clustering,90blake2012wigglez,87scolnic2018complete,86yu2018hubble}.

The cosmological constant $\Lambda$ is presented to supply the simplest illustration for the late time acceleration within the framework of general relativity (GR), associated in the current $\Lambda$ CDM model.The adjusting problem which effects from the important difference between observed and theoretical values of $\Lambda$ and the cosmic coincidence problem, which interrogate, why the energy densities of matter and dark energy are of the same ordering in the usual period, are the two primary theoretical issues with model, in spite of the reality that it effectively explains the plurality of cosmological observations \cite{30perlmutter2003measuring,31caldwell2004cosmic,70zlatev1999quintessence}.

Scalar field based vigorous models have pictured a lot of interest between the indicated substitutes for the cosmological constant. The observed acceleration can be brought dynamically by the concept of \textit{quintessence}, which is driven by a cosmological scalar field that is slightly associated to gravity \cite{68ratra1988cosmological,69copeland2006dynamics,71steinhardt1999cosmological}. In order to account for more complicated evolutions, extensions like phantom models  ($w<-1$) and $k$-essence \cite{65caldwell2002phantom,73armendariz2000dynamical,78chiba2000kinetically,37matsumoto2010reconstruction} involve non canonical kinetic terms or potential.

The resultant \textit{quintom} cosmologies clarify transitions amongst accelerating and decelerating aspects by allowing the equation of state parameters to betray the cosmological constant boundary ($w = -1$) among a combination of phantom and quintessence fields \cite{66setare2009quintom,79sadjadi2006transition,99zhao2005perturbations}. These models can unite the inflationary and dark energy periods into a single structure in addition to illustrating late time cosmic acceleration \cite{83barrow1988string,84barrow1990graduated,36nojiri2006unifying}.The exploration of gravitational alternative as a geometric origin  of dark energy is eager by the fact that such scalar field models are delicate to initial provisions and often suffer from adjusting problems.

Varying geometric sector of Einstein's field equations gives a possible substitute for DDE models. The Einstein–Hilbert action is generic to a nonlinear function of the Ricci scalar R in the f(R) gravity theory, one of the most straight and highly studied extensions \cite{53capozziello2002curvature,54carroll2004cosmic,55capozziello2006dark,56bohmer2008dark,52sotiriou2010f}. The late time acceleration can be naturally illustrated by these models without the demand for exotic matter field. Though it was found out that early formulation such as $f(R)=R-\mu^4/R$ were inconsistent with stability requirements and solar system tests \cite{57chiba20031,58erickcek2006solar}.

Higher order curvature modifications were introduced to conquer these boundaries, consequent in theories like $f(G)$ gravity, which is based on the Gauss Bonnet Invariant $G$ \cite{42nojiri2005modified,43de2009construction,44cognola2006dark,47bamba2010finite};$f(T,B)$ gravity, which holds boundary terms and torsion \cite{48bahamonde2018thermodynamics}; and $f(Q)$ gravity, which is designed from non-metricity scalar $Q$ \cite{91koussour2023new}. Both early time inflation and late time acceleration may be depicted by these generalized models using a single geometric framework \cite{46elizalde2007stationary,40bengochea2009dark,41myrzakulov2012gravity}.

Myrzakulov gave a beyond theory by presenting $f(R,T)$ gravity in which the gravitational Lagrangian rests on the trace of the energy momentum tensor $T$ and the Ricci scalar $R$  \cite{59myrzakulov2012frw}. Certainly in the lack of a cosmological constant, the efficient interaction made by this non-minimal coupling amongst matter and geometry can render cosmic acceleration. This theory has been extended in a different ways for which anisotropic and bulk viscous models are between cosmological contexts including $f(R,T)=f_1(R)+f_2(R)f_3(T)$ \cite{60bhardwaj2020some,61yadav2020existence,62sharma2020power,63sharma2022scalar,97bhardwaj2018non,98bhardwaj2019lrs}.

These developments created by Singh \textit{et al.}  \cite{82singh2023cosmological} to pose the $f(R,T^\psi)$ gravity framework, in which the gravitational action rests on the both $T^\psi$ and the Ricci scalar $R$ This describes a connection between a scalar field $\psi$ and the matter sector (via its trace $T$). By giving a dynamic interplay among geometry, matter and scalar field cosmology. A rich cosmological dynamics that can explain transformations amongst quintessence and phantom regiments is  made possible by the introduction of scalar field \cite{77jawad2015correspondence,80shamir2020f,81malik2020noether}. Besides, with the precise parameters selection, the theory can meet the fundamental energy necessities, assuring its physical feasibility \cite{92santos2005energy,93santos2007energy,95capozziello2018role,96bergliaffa2006constraining}. As a result, $f(R,T^\psi)$ gravity gives more understanding structure for investigating the part that matter geometry connections and scalar field play in the cosmic evolution.

Such modified gravity theories have a vast chance to be tested in the increasingly correct modern cosmological data. The restrictions on a cosmological parameters are provided by the very latest Pantheon+ compilation of type la- Supernovae \cite{87scolnic2018complete}, BAO data from the SDSS and Wigglez audits \cite{88beutler20116df,89anderson2014clustering,90blake2012wigglez}, and CMB anisotropy measurements from Planck \cite{101ade2016planck}. Additional traits for distinguishing between competing cosmological models are delivered by studies using the hubble parameter $H(z)$ and deceleration parameter $q(z)$  \cite{85xu2008constraints,86yu2018hubble,100sahni2008two}.

A credible framework that can replicate the observed late time acceleration, coherent with energy conditions and possibly give coherent explanation of the universe's transition from deceleration to acceleration is $f(R,T^\psi)$ gravity inside this observational landscape.

Density and analyzing a cosmological model within the framework of $f(R,T^\psi)$ gravity taking into chronology and latest observational constraints like $H(z)$, BAO and Pantheon datasets is the aim of the ongoing study. We examine however this theory can explain the current cosmic acceleration, the deceleration- acceleration transformation  and the accomplishment of energy conditions by embracing appropriate functional forms of $f(R,T^\psi)$. Additionally, study explores how a better comprehension of dark energy and the developments and of the universe can be gained through the interaction of curvature, matter and scalar field dynamics in $f(R,T^\psi)$ gravity.

\section{Field equations and first order formalism}
		\label{sec2}

The action for $f(R, T^{\psi})$ gravity is given by

\begin{equation}
S = \int d^{4}x \, \sqrt{-g} \left[ f(R, T^{\psi}) + \mathcal{L}(\psi, \partial_{\nu}\psi) \right],
\end{equation}
Assume that $16\pi G = c = 1$.$f(R,T)$  be an explicit function in $R$ and $T$. Accordingly, braneworld scenario, it found to be stable ~\cite{23bazeia2015thick}. If $f(R,T)$ is linear in $R$, then solutions tends the FRW model in hig red-shift governance ~\cite{24baffou2015cosmological}. According to linear and explicit form of $f(R,T^{\psi})$~\cite{22moraes2016complete}.

\[
f(R, T^{\psi}) = -\frac{R}{4} + \mu T^{\psi},
\]
where $\mu$ is a constant. The corresponding field equations are

\begin{equation}
G_{ij} = 2 \left[ T^{\psi}_{ij} - g_{ij} \mu T^{\psi} - 2\mu \, \partial_{i}\psi \, \partial_{j}\psi \right],
\end{equation}
where $G_{ij}$ and $T^{\psi}_{ij}$ denotes Einstein tensor and Energy momentum tensor of a scalar field. The Lagrangian density and the energy-momentum tensor for real scalar field $\psi$ are given by

\begin{equation}
\mathcal{L} = -\frac{1}{2} \, \partial_{i}\psi \, \partial^{i}\psi - V(\psi),
\end{equation}

\begin{equation}
T^{\psi}_{ij} = \partial_{i}\psi \, \partial_{j}\psi - g_{ij}\mathcal{L},
\end{equation}
where $V(\psi)$ is the self-interacting potential. The trace of the energy-momentum tensor is

\begin{equation}
T^{\psi} = \dot{\psi}^{2} + 4V(\psi),
\end{equation}
where the dot shows derivative with respect to $t$.  The line element of the LRS Bianchi type-I universe model is given by

\begin{equation}
ds^{2} = -dt^{2} + X^{2}(t)\,dx^{2} + Y^{2}(t)\,(dy^{2} + dz^{2}).
\end{equation}
Shear and expansion scalar for this metric are as follows

\begin{equation}
\sigma^{2} = \frac{1}{3}\left( \frac{\dot{X}}{X} - \frac{\dot{Y}}{Y} \right)^{2},
\qquad
\Theta = \frac{\dot{X}}{X} + 2\frac{\dot{Y}}{Y}.
\end{equation}
Assume that both shear and expansion scalar are proportional to each other 
each other ($\theta \propto \sigma$), which steers to the relation $X = Y^{n}$, where $n \neq 0$ is a constant~\cite{25shamir2015locally}. The mean Hubble parameter is given as

\begin{equation}
H = \frac{1}{3}\left( \frac{n + 2}{n} \right)\frac{\dot{X}}{X}.
\end{equation}
The correspondent field equation become
\begin{equation}
\frac{9}{2} \frac{(2n + 1)}{(n + 2)^{2}} H^{2}
= \left( \frac{1}{2} - 2\mu \right) \dot{\psi}^{2} + \mu T^{\psi} - V(\psi),
\end{equation}

\begin{equation}
\frac{3\dot{H}}{(n + 2)} + \frac{3}{2} \frac{(3H)^{2}}{(n + 2)^{2}}
= -\left( \frac{1}{2} \dot{\psi}^{2} - \mu T^{\psi} \right) - V(\psi),
\end{equation}
\begin{equation}
\frac{3}{2}\left(1 + \frac{1}{n}\right)\frac{\dot{H} n}{(n + 2)} 
+ \frac{9}{2}\left(1 + \frac{1}{n} + \frac{1}{n^{2}}\right)\frac{H^{2} n^{2}}{(n + 2)^{2}}
= -\left( \frac{1}{2} \dot{\psi}^{2} - \mu T^{\psi} \right) - V(\psi).
\end{equation}
The anisotropy parameter is defined as~\cite{20sharif2012anisotropic}

\begin{equation}
\mathcal{A}_{p} = \frac{1}{3} \sum_{i=1}^{3} \left( \frac{\Delta H_{i}}{H} \right)^{2} \; ; \qquad
\Delta H_{i} = H - H_{i},
\end{equation}
where $H_i$ stands for the directional Hubble parameters. In our case, it becomes

\begin{equation}
\mathcal{A}_{p} = \frac{2(n - 1)^{2}}{(n + 2)^{2}}.
\end{equation}
It can be noticed that the anisotropy parameter decreases for $-2 < n \leq 1$, while it increases for $-\infty < n < -2$ and $1 \leq n < \infty$. For the scalar field, the equation of motion is given by

\begin{equation}
(1-2\mu)\big(\ddot{\psi}+3H\dot{\psi}\big) + (1-4\mu)\,V_{\psi} = 0
\end{equation}

where the subscript $\psi$ denotes derivative with respect to $\psi$. 
Bazeia \textit{et al.}~\cite{8bazeia2006first} alleged the first-order formalism based on the assumption $H = W(\psi)$ to solve the field equations. We relate this 
formalism and find the values of $H$, $\omega_{\text{eff}}$, and $V(\psi)$. Equations~(8) and~(9) yield

\begin{equation}
\dot{H} = \frac{(n + 2)}{(2n + 1)} 
\left[ -\frac{(n + 2)}{3} + 2\mu \right] \dot{\psi}^{2}
+ \frac{2(n - 1)(n + 2)}{3(2n + 1)} \left( \mu T^{\psi} - V \right).
\end{equation}
By using assumption of first order formalism, the above equation (15) becomes

\begin{equation}
\frac{(n + 2)(2\mu - 1)}{3}\dot{\psi}^{2}
- W_{\psi}\dot{\psi}
+ \frac{3(n - 1)}{(n + 2)}W^{2}.
\end{equation}
From equation (8), the potential of the scalar field can be written as follows

\begin{equation}
V(\psi) = \frac{1}{1 + 4\mu}
\left[
\left( \frac{1}{2} - \mu \right)\dot{\psi}^{2}
- \frac{9}{2} \frac{(2n + 1)}{(n + 2)^{2}} W^{2}
\right].
\end{equation}
The Equation of State parameter is estimated as

\begin{equation}
\omega_{\mathrm{eff}} = \frac{p_{\mathrm{eff}}}{\rho_{\mathrm{eff}}}
= 1 + \frac{2(n + 2)^{2}\left(2\mu - \tfrac{1}{4}\right)\dot{\psi}^{2}}
{9(2n + 1)W^{2}}.
\end{equation}
Now, we account $W(\psi)$ in the form of exponential, polynomial, 
as well as trigonometric functions and solve Eq.~(15) for $\psi(t)$. We first take $W(\psi)$ as an exponential function given by~\cite{8bazeia2006first}

$$W(\psi) = e^{a_{1}\psi} \quad \Rightarrow \quad W_{\psi} = a_{1} e^{a_{1}\psi}$$
where $b_1$ is real constant. The scalar field potentials succeeded by value of $W(\psi)$. When we take $b_1=1$ it forms some negative potential and $b_1=2$ directs to potential which shows spontaneous symmetry breaking ~\cite{8bazeia2006first}. We examine actions of potential for this model in $f(R, T^{\psi})$ theory. Substituting $W$ and $W_{\psi}$ in Eq. (15), it follows that

\begin{equation}
\psi(t) = -\frac{1}{b}
\ln \left[
    -b_{1}c_{1} + \frac{3b_{1}t}{2(n + 2)(2\mu - 1)}
\right]
\times
\left\{
    -b_{1} \pm \sqrt{b_{1}^{2} - 4(2\mu - 1)(m - 1)}
\right\}.
\end{equation}
Here $c_1$ is the constant of integration. This gives two values of $\psi(t)$, which are indicated by $\psi_{-}$ and $\psi_{+}$ for negative and positive signs, respectively. After substituting value of $\psi$, the analogous values of $H$, $\omega_{\text{eff}}$, and $V(\psi)$ become

\begin{equation}
H = 
\left[
    -b_{1}c_{1} + \frac{3b_{1}t}{2(n + 2)(2\mu - 1)}
    \left( -b_{1} \pm \sqrt{b_{1}^{2} - 4(2\mu - 1)(n - 1)} \right)
    (n - 1)^{\frac{1}{2}}
\right]^{-1}
\end{equation}
but, $$t(z)=\frac{1}{m\alpha}log(1+(1+z)^{-m})$$

\begin{equation}
H(z) = 
\left[
    -b_{1}c_{1} + \frac{3b_{1}log(1+(1+z)^{-m})}{2m\alpha(n + 2)(2\mu - 1)}
    \left( -b_{1} \pm \sqrt{b_{1}^{2} - 4(2\mu - 1)(n - 1)} \right)
    (n - 1)^{\frac{1}{2}}
\right]^{-1},
\end{equation}

\begin{equation}
\omega_{\mathrm{eff}} = 
1 + 
\frac{\left(-\tfrac{1}{4} + 2\mu\right)}
{18(2n + 1)(2\mu - 1)^{2}}
\left(
    -b_{1} \pm \sqrt{b_{1}^{2} - 4(2\mu - 1)(n - 1)}
\right)^{2}
(n - 1).
\end{equation}

\begin{equation}
\begin{aligned}
V(\psi) &= \Bigg[ 9\left( \frac{1}{2} - \mu \right)
\left( -b_{1} \pm \sqrt{b_{1}^{2} - 4(2\mu - 1)(n - 1)} \right)^{2}
- 18(2n + 1)(2\mu - 1)^{2} \Bigg] \\
&\quad \times \Bigg[ (1 + 4\mu)
\Big\{ -2b_{1}c_{1}(n + 2)(2\mu - 1)
+ 3b_{1}t\left( -b_{1} \pm \sqrt{b_{1}^{2} - 4(2\mu - 1)(n - 1)} \right) \Big\}^{2} \Bigg]^{-1}.
\end{aligned}
\end{equation}

\begin{figure}[H]
			\centering
			\includegraphics[width=0.8\textwidth]{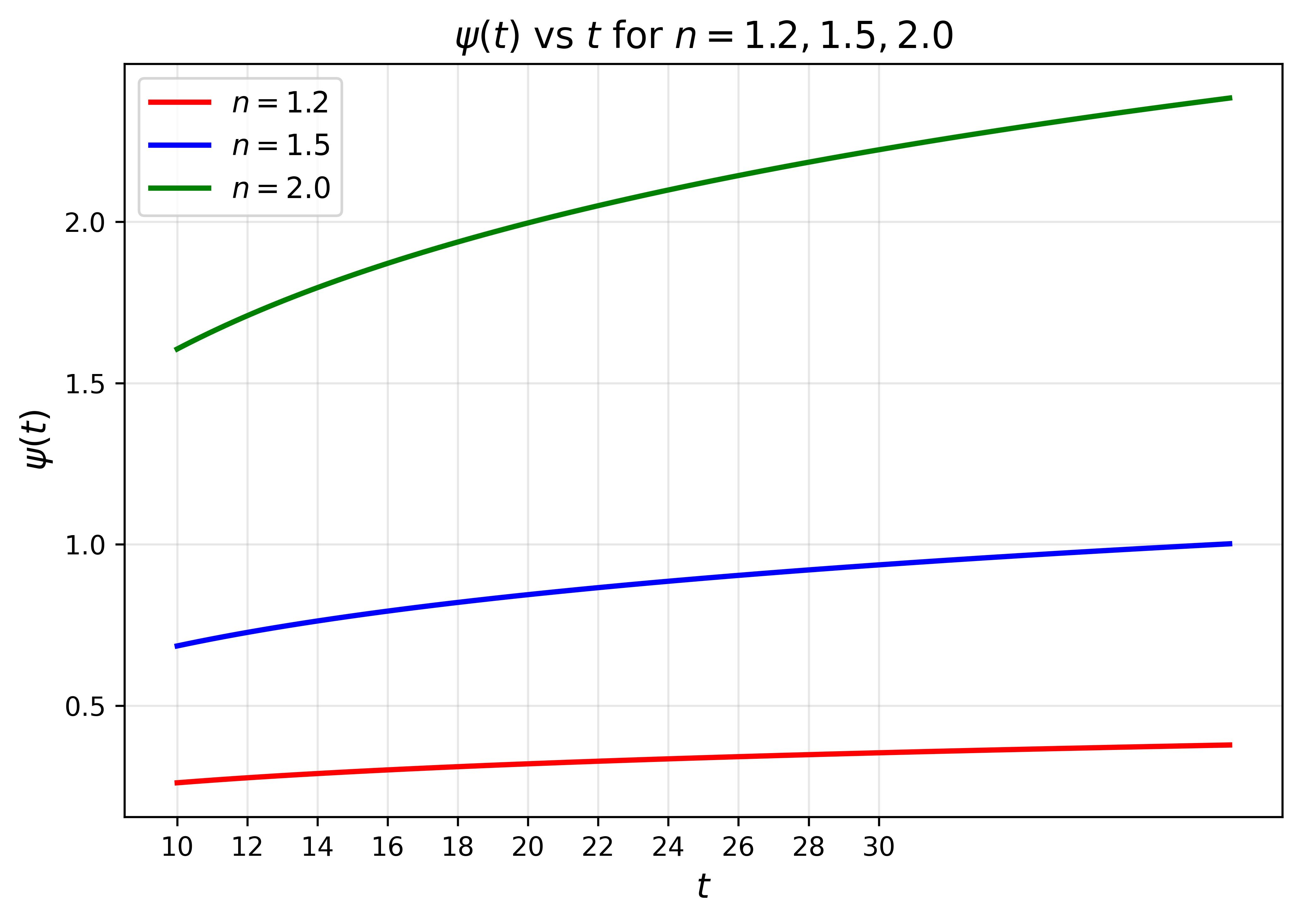}
			\caption{Evolution of the scalar-field $\psi$ as a function of cosmic time t for different values of the anisotropy parameter n.}\label{1}
		\end{figure}

The Figure~\ref{1} indicates the progression of the scalar field $\psi(t)$ with cosmic time $t$ for three different values of the model parameter $n$. For all three curves, $\psi(t)$ increases monotonically as time processes, showing a steady growth of the scalar field in this modified--gravity set. The parameter $n$ strongly weights both the magnitude and the growth rate of $\psi(t)$: the curve with the smallest value, $n = 1.2$ (red), has the lowest amplitude and the slowest rise, while larger values of $n$ lead to noticeably higher values of the scalar field. The curve for $n = 1.5$ (blue) lies above the red curve throughout the evolution, and the curve for $n = 2.0$ (green) indicates the fastest and largest growth, passing values above $4$ at $t = 100$. The separation between the curves increases with time, showing that the parameter $n$ supplements the evolution of $\psi(t)$ more vigorously at late times. Overall, the plot establishes that increasing $n$ improves both the initial value and the growth rate of the scalar field $\psi(t)$.

\begin{figure}[H]
    \centering
    \includegraphics[width=0.8\textwidth]{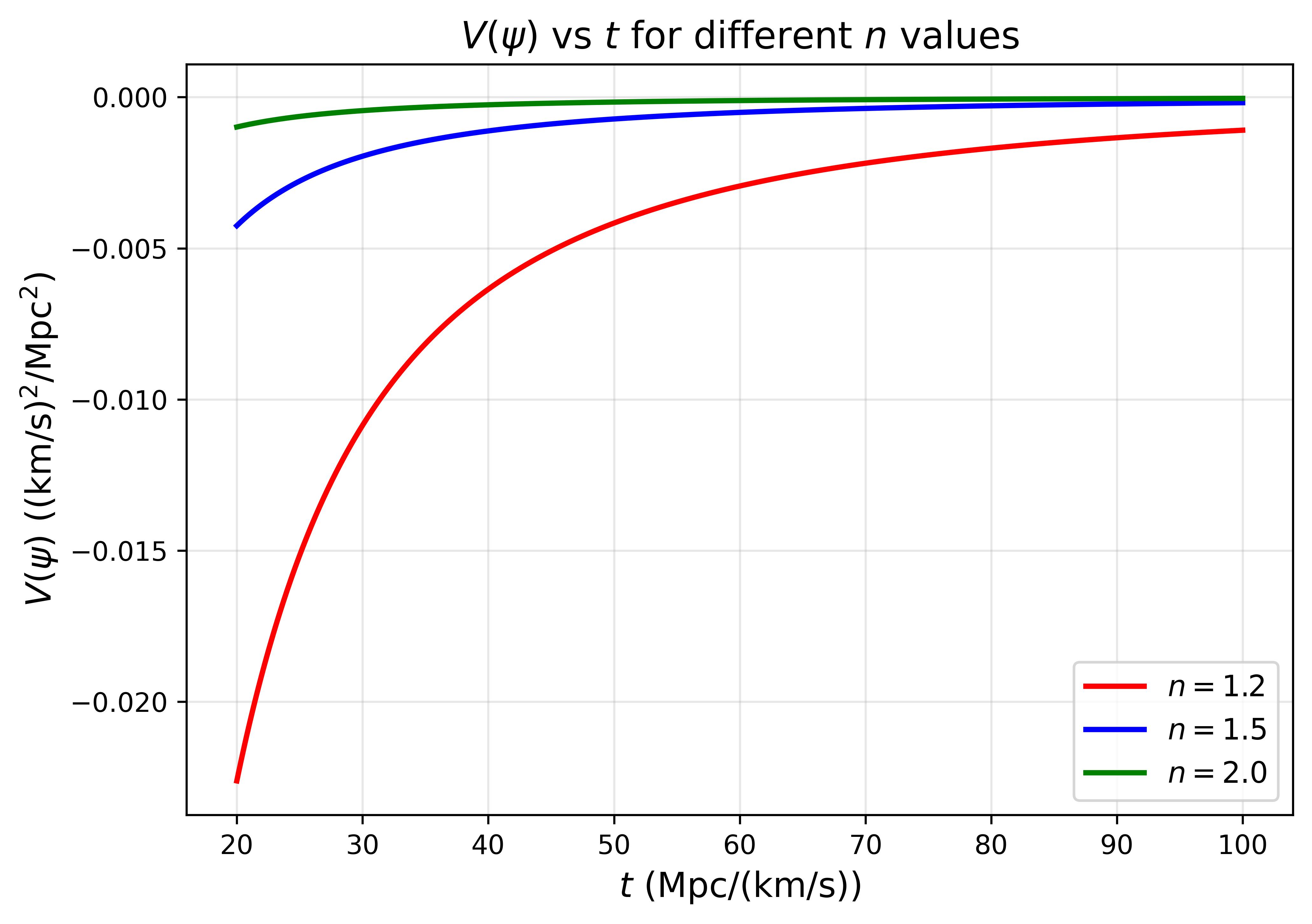} 
    \caption{Evolution of the scalar-field potential V($\psi$)as a function of cosmic time t for different values of the anisotropy parameter n.}
     \label{2}
\end{figure}

The Figure~\ref{2} explains the versions of the potential function $V(\psi)$ with respect to cosmic time $t$ for diverse values of model parameter $n$, as obtained from Equation~(23). The curves corresponds to $n = 1.2$ (red), $n = 1.5$ (blue), and $n = 2.0$ (green). It is possible that the potential $V(\psi)$ displays a negative behavior across the considered range. If $t$ increases, it progressively coming to zero. This shows that the potential becomes smooth at later cosmic times, denoting a slower rate of change in scalar field dynamics. Besides, higher values of $n$ submit less negative potentials, indicating that strength of the potential decreases with increasing $n$. Such behavior reproduces the importance of the parameter $n$ on the evolution of the scalar field and the overall dynamics of the cosmological model constrained by Equation~(23).

\section{Observational data analysis}
		\label{sec3}

To know the best fit values of model parameters, Metropolis--Hastings procedure-based Markov Chain Monte Carlo (MCMC) mechanism has been used in this section. For any noticeable physiologic quantity $\zeta$, the theoretically expected value is represented by $\zeta_{\text{th}}$, and the analogous observational value is signified by $\zeta_{\text{ob}}$.The $\chi^{2}$ function for hubble data is defined as \cite{dhankar2025observational}

\begin{equation}
\chi^{2}_{\zeta}(P) = \sum_{i=1}^{N} 
\frac{\left[\zeta_{\text{th}}(P) - \zeta_{\text{ob}}\right]^{2}}
{\sigma_{\zeta}^{2}}.
\end{equation}
where $\sigma_{\zeta}$ is standard deviation in observations of a physical quantity, and $P$ denotes for the model parameters.

In the spatially flat spacetime, the distance modulus for Pantheon is defined as \cite{chang2019constraining}
\begin{equation}
    \mu_{\rm th} = 5 \log_{10} \left( \frac{d_L}{\rm Mpc} \right) + 25,
\end{equation}
where the luminosity distance is given as $d_L = (c/H_0) D_L$, 
$H_0$ is the Hubble constant, $c$ is the speed of light and $D_L$ 
takes the form,
\begin{equation}
    D_L = (1+z_{\rm cmb}) \int_0^{z_{\rm cmb}} \frac{dz}{E(z)},
\end{equation}
where $z_{\rm cmb}$ denotes the CMB frame redshift. The expression 
for $E(z)$ varies in different cosmological models.

The $\chi^{2}$ function of the (SNIa) measurements is given by \cite{dhankar2025observational}.

\begin{equation}
\chi^{2}_{SN}(\phi^{\nu}_{s}) = \mu_{s}\, C^{-1}_{s,\mathrm{cov}}\, \mu^{Transpose}_{s},
\end{equation}

where

$$\mu_{s} = \{\mu_{1} - \mu_{\mathrm{th}}(z_{1},\phi^{\nu}),\; \ldots,\;
\mu_{N} - \mu_{\mathrm{th}}(z_{N},\phi^{\nu})\}.$$

The extremely probable values of the model parameters can be establish by statistically minimizing the assessment function $\chi^{2}$. We have practiced observational data from the Pantheon compilation, which contains Pantheon 1048 Type la Supernovae (SN la) possible magnitudes in the redshift range $0.01 \leq z \leq 2.26$, observations datasets of baryon acoustic oscillation (BAO) 17 points, and observational Hubble data (OHD) having 30 points restricted in the range $0.106 < z < 2.340$ and $0.07 < z < 2.0$ of redshift respectively. There are two dimensional confidence contour plots and one dimensional marginal plots for the model parameters at the $1\sigma$ (68\%) and $2\sigma$ (95\%) confidence levels of the provided model.  Table~1 reprises the best fit (or best-approximated) parameter values of the developed model.

\begin{figure}[H]
    \centering
    \includegraphics[width=0.8\textwidth]{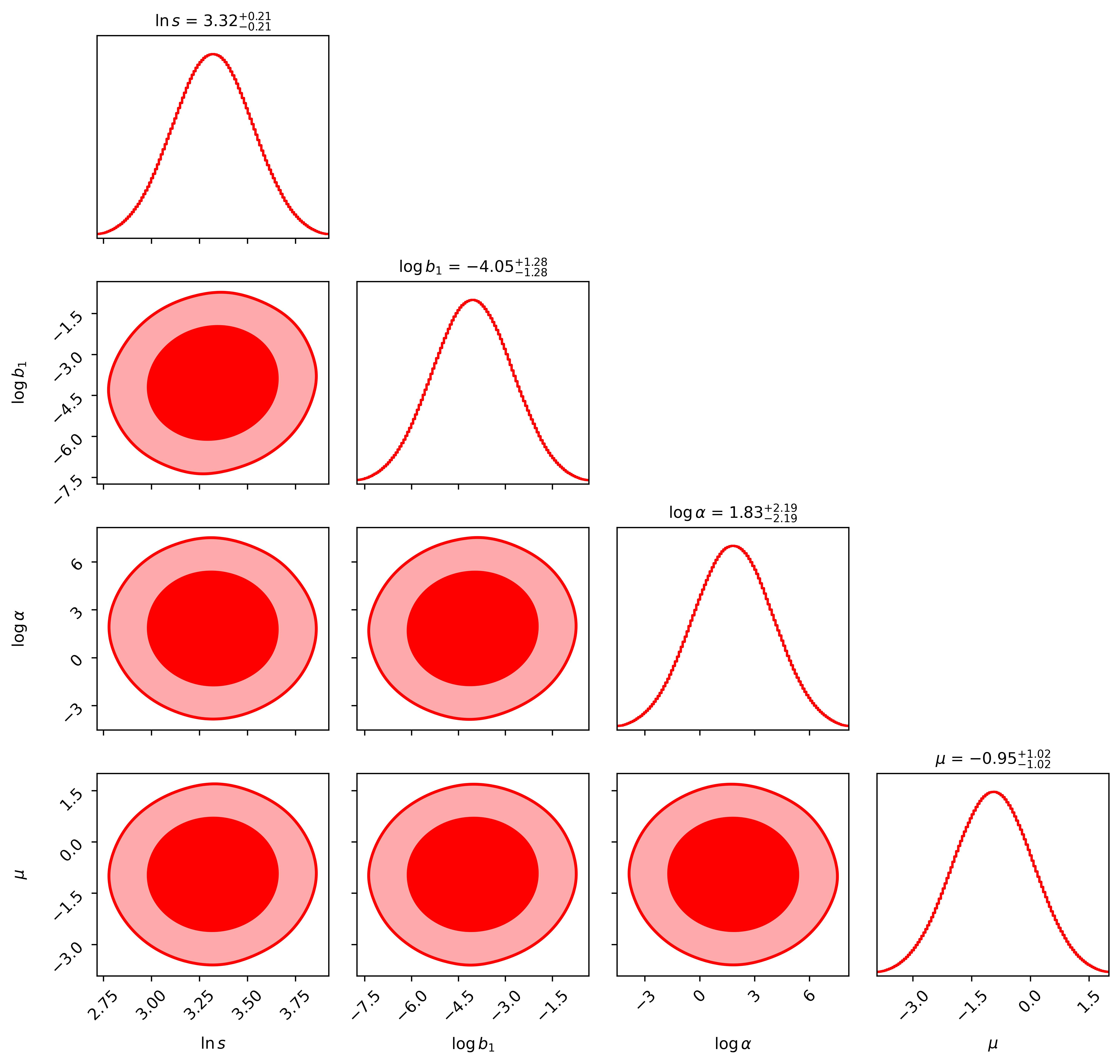} 
    \caption{Confidence contours and marginalized posterior distributions for the parameters $\ln s$, $\log b_{1}$, $\log \alpha$, and $\mu$ obtained from the Hubble (OHD) dataset. }
     \label{3}
\end{figure}

The combined allocations of the model parameter $\ln s$, $\log b_{1}$, $\log \alpha$, and $\mu$ developed using a Markov Chain Monte Carlo (MCMC) inspection are shown by the corner plot in Figure~\ref{3}. The one dimensional marginalization probability allocations for each parameter are shown in diagonal panels simultaneously, with their $68\%$ confidence intervals. The two dimensional relationship contours between couples of parameters are described in off diagonal panels; The plot denotes two confidence levels contours analogous to $1\sigma$ ( $68\%$ confidence level) and $2\sigma$ ($95\%$ confidence level), where the inner red area shows the $1\sigma$ uncertainty and outer light red shaded region shows $2\sigma$ uncertainty. Due to Gaussian distributions, the parameters are highlight to be fine limitation, indicating true convergence and accurately evaluated. Efficient sampling is confirmed by MCMC chains with mean acceptance fraction of 0.177, the model selection statistics shows that $\mathrm{AIC}_{\text{model}} = 268.065$ and $\mathrm{BIC}_{\text{model}} = 277.873$. An awesome consent between the model and the observational data is implicit by the reduced chi-square $\chi^{2}_{\text{red}} = 0.3074$, which has the associated chi-square value at the MAP point, $\chi^{2} = 7.3786$ .

\begin{figure}[H]
    \centering
    \includegraphics[width=0.8\textwidth]{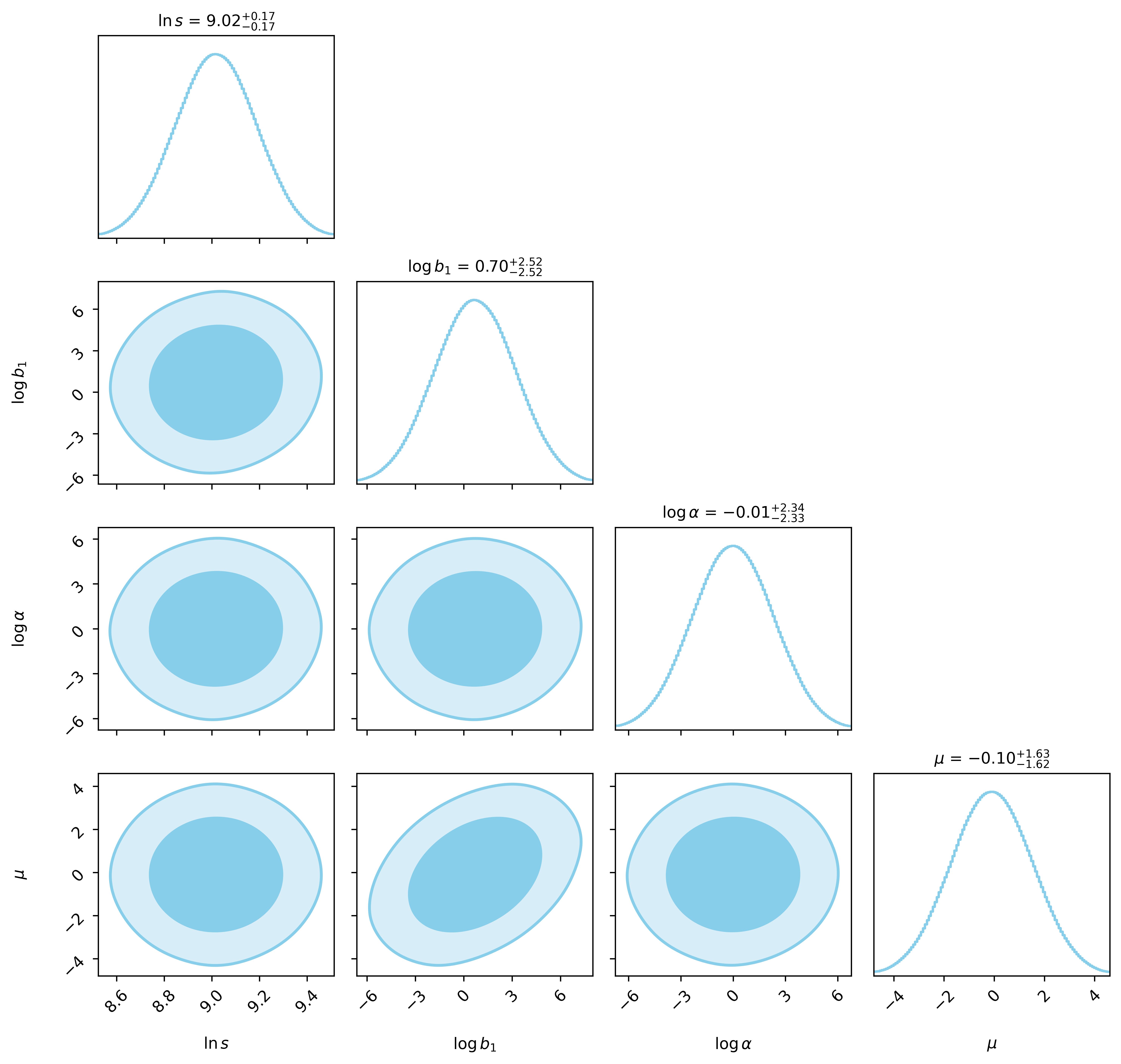} 
    \caption{Joint posterior distributions and confidence contours for the model parameters obtained from BAO data.}
     \label{4}
\end{figure}

Applying a Markov Chain Monte Carlo (MCMC) investigation, the combined posterior distributions of the model parameters $\ln s$, $\log b_{1}$, $\log \alpha$, and $\mu$ are shown in the corner plot in Figure~\ref{4}. The one dimensional marginalization probability allocations for each parameter are shown in diagonal panels simultaneously with their $68\%$ confidence interval. The plota shows two confidence regions contours corresponding to $1\sigma$ ($68\%$ confidence level) and $2\sigma$ ($95\%$ confidence level), where inner blue area shows the $1\sigma$ uncertainity and outer light blue area shows $2\sigma$ uncertainity. Here 0.428 is the mean acceptance fraction shows effective sampling and parameters shows nearly Gaussian posteriors and look well restricted, indicating a reliable convergence of MCMC chain. The model selection statistics denotes $\mathrm{AIC}_{\text{model}} = 339.346$ and $\mathrm{BIC}_{\text{model}} = 345.179$. With $11$ degrees of freedom, the best fit best-fit chi-square value is $\chi^{2} = 3.4252$, resultant in a reduced chi-square $\chi^{2}_{\text{red}} = 0.3114$. This shows that the model and observational data have great consensus. In general, the findings indicates that the model gives a true and consonant chronology of the cosmological data and that decided parameter behaves fine.

\begin{figure}[H]
    \centering
    \includegraphics[width=0.8\textwidth]{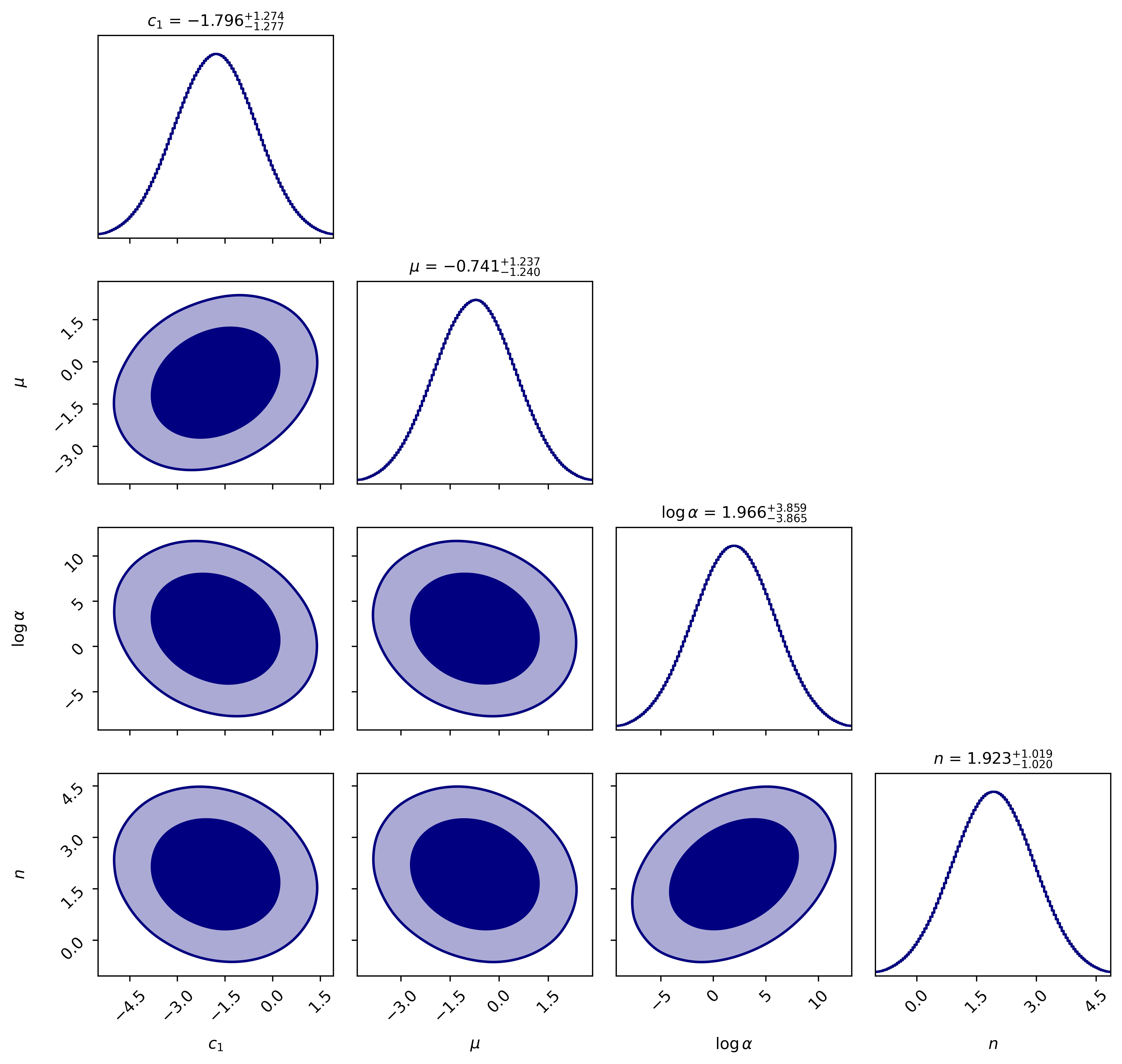} 
    \caption{MCMC-based posterior distributions for parameters $c_1$, $\mu$, $\log \alpha$, and $n$ using the Pantheon Type Ia Supernova dataset.}
     \label{5}
\end{figure}

The corner plot of the cosmological model parameters decided from the Pantheon Supernova dataset is displayed in Figure~\ref{5}. This figure shows two dimensional confidence level contours (68\% and 95\%) as well as one dimensional marginalization's probability allocations along the diagonals for the parameters $c_1$, $\mu$, $\log \alpha$, and $n$. The model's strength is further endorsed by the statistical results, which highlights an excellent fit to the Pantheon data with $\mathrm{AIC} = 3078.31$ and $\mathrm{BIC} = 3112.98$, along with reduced chi-square of $\chi^2_\nu = 0.98$ and $\chi^2 = 1019.2$ for 1040 degrees of freedom. These statistical values also features the stability and consistency of the model, verifying its capacity to accurately depict the late time acceleration.

\begin{figure}[H]
    \centering
    \includegraphics[width=0.8\textwidth]{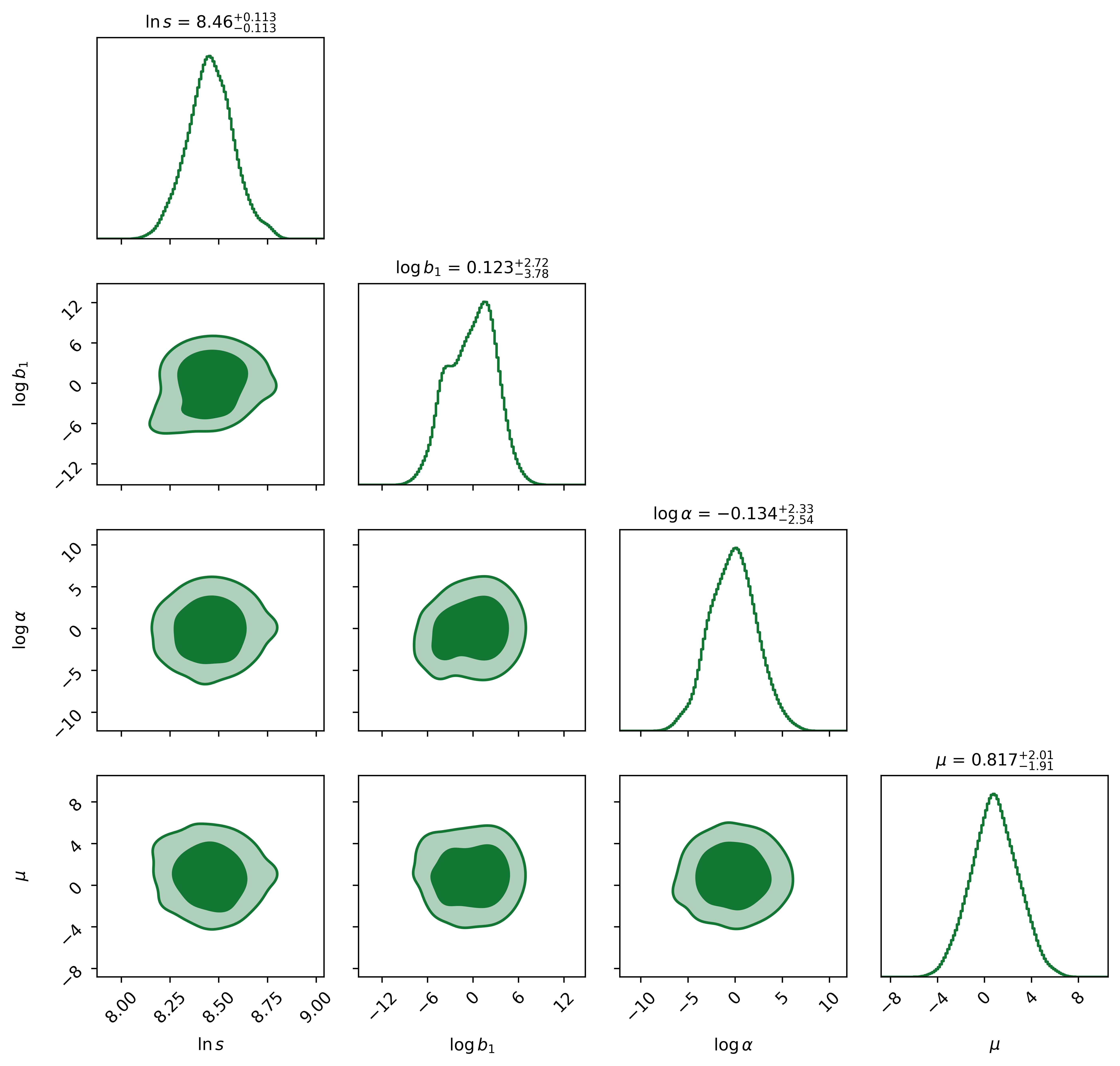} 
    \caption{Combined parameter constraints from joint Hubble (OHD) and BAO datasets.}
     \label{6}
\end{figure}

A joint fit to Hubble parameter ($H$) and Baryon Acoustic Oscillation (BAO) data submitted the combined posterior allocations and marginalization one-dimensional (1D) likelihoods of model parameters $\ln s$, $\log b_1$, $\log \alpha$, and $\mu$, which are shown in the Figure~\ref{6}. The diagonal forum shows the marginalized 1D posterior allocations with best fit values and $1\sigma$ uncertainty, since the contours of the two dimensional (2D) plots described the 68\% and 95\% confidence regions, showing parameter relationships. With a total 47 data points ($N_H = 30$, $N_{BAO} = 17$) and $40$ degrees of freedom, the combined fit statistics disclose a good model consistency with the data.  The reduced chi-square value $\chi^2_{\text{red}} = 0.095$ indicates an excellent fit. The models sufficiency is beyond advocated by the Bayesian Information Criterion (BIC $= 892.589$) and the Akaike Information Criterion (AIC $= 879.638$).

\begin{figure}[H]
    \centering
    \includegraphics[width=0.8\textwidth]{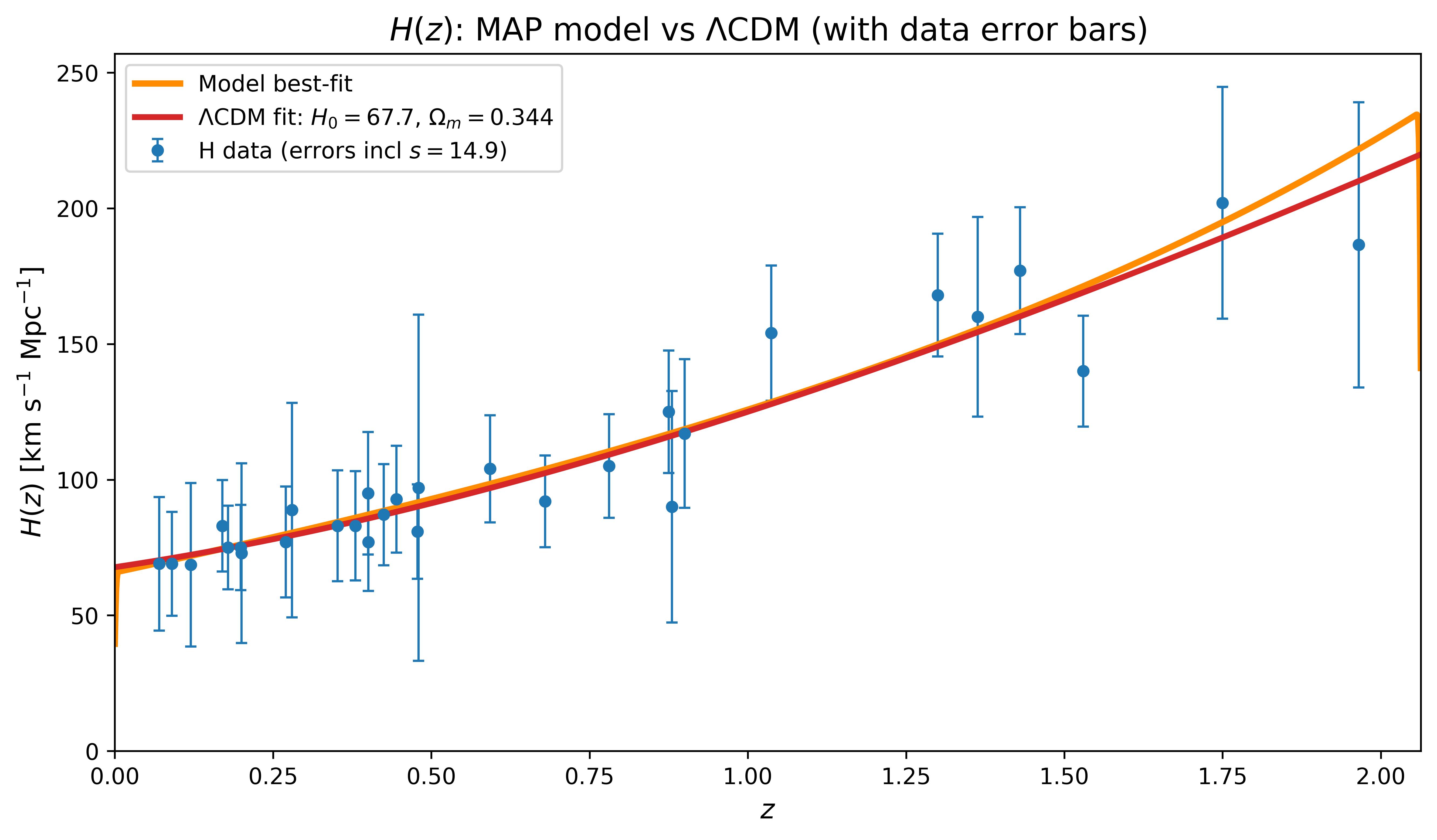} 
    \caption{Comparison of the reconstructed Hubble function H(z)(orange curve) with observational H(z) data (blue points) and the $\Lambda CDM$ prediction (red curve).}
     \label{7}
\end{figure}

Applying observational $H(z)$ data and associated error bases, Figure~\ref{7} compares the conventional $\Lambda$ CDM cosmology with the rebuilt Hubble parameter $H(z)$ from the maximum a posteriori (MAP) model. The red curve shows the $\Lambda$ CDM fit, which is defined by $H_0 = 67.7\, \mathrm{km\, s^{-1}\, Mpc^{-1}}$, and $\Omega_m = 0.344$. The orange curves denotes the best fit prophecy of the MAP model, while the blue points indicates the observed Hubble parameter values at different redshifts, including their uncertainties. The MAP model indicates a barely larger expansion rate at higher redshifts, but it nearly matches the observational data and remains coherent with the  $\Lambda$ CDM directed at lower redshifts. This consistency inside observational errors mounts the indicated models harmony with current cosmological observations by showing that it serves a reliable and accurate account of the cosmic expansion history.

\begin{figure}[H]
    \centering
    \includegraphics[width=0.8\textwidth]{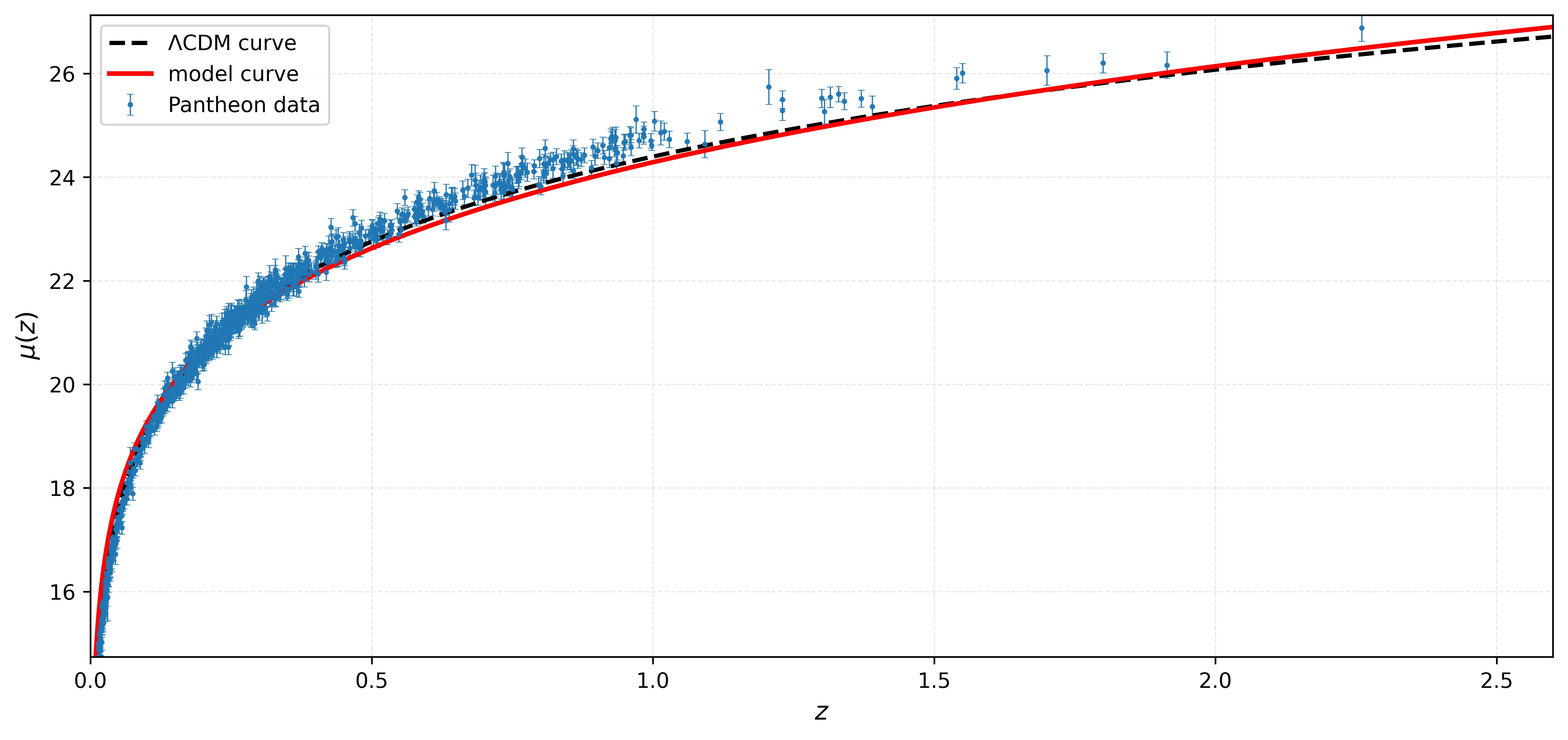} 
    \caption{Comparison between the theoretical distance modulus $\mu(z)$ predicted by the reconstructed model (solid red line) and the Pantheon Supernova observations (blue data with error bars).}
     \label{8}
\end{figure}

The Figure~\ref{8} compares the observational data and Pantheon Type la Supernovae ( blue points with error bars ) with the theoretical distance modulus $\mu(z)$ read by the rebuilt cosmological model ( Solid red line ). For visual analogy, the average $\Lambda$ CDM model is described by the black dashed curve, which has been moved to match with data. Above the whole redshift range ($0 < z < 2.5$) , the rebuilt model and the observational data overlap nearly, indicating a good match and high degree of consistency with observational measurements. The average continuity of roughly 0.1264 may reveals minimal variation between the model and $\Lambda$ CDM, displaying that the indicated model successfully imitates the observed cosmic acceleration while keeping compliancy with the mainstream cosmological framework.

\begin{figure}[H]
    \centering
    \includegraphics[width=0.8\textwidth]{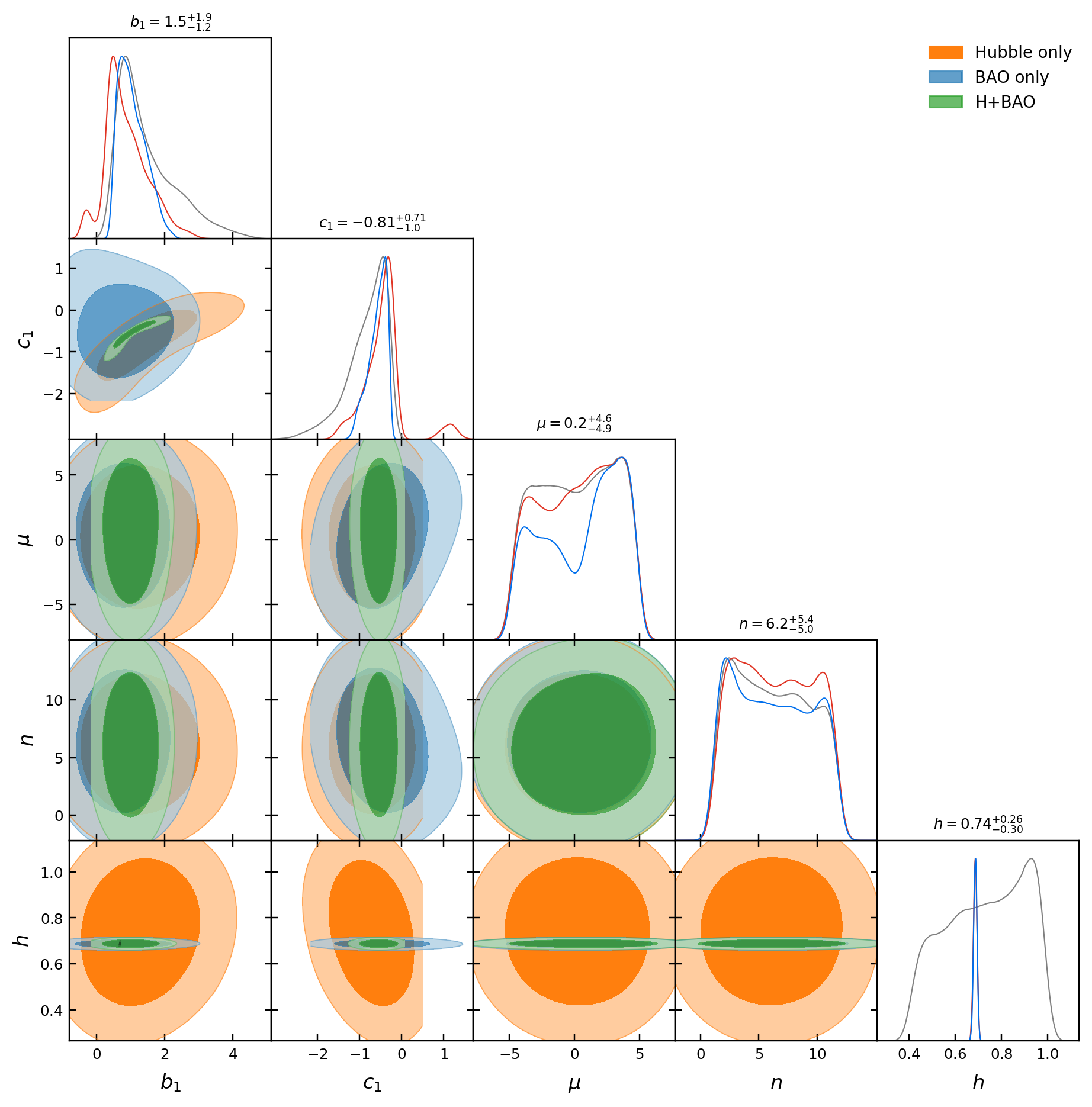} 
    \caption{Joint posterior distributions for parameters $b_1$, $c_1$, $\mu$, $n$, and $h$ obtained using Hubble-only, BAO-only, and combined Hubble+BAO datasets.}
     \label{9}
\end{figure}

Figure~\ref{9} gives corner plot explaining the joint posterior allocations and marginalized one-dimensional probability densities for the model having parameters $b_1$, $c_1$, $\mu$, $n$, and $h$, deduced from three distinct datasets : \emph{Hubble only} (orange), \emph{BAO only} (blue), and the combined \emph{Hubble + BAO} (green). The best-fit values over with their analogous $1\sigma$ (68\% confidence level) and $2\sigma$ (95\% confidence level) uncertainties are obtained as $b_1 = 1.5^{+1.9}_{-1.0}$, $c_1 = -0.81^{+0.71}_{-0.91}$, $\mu = 0.2^{+4.8}_{-4.5}$, $n = 6.2^{+3.6}_{-5.4}$, and $h = 0.74^{+0.38}_{-0.36}$.The internal, darker contours shows the $1\sigma$ (68\% confidence area), while the outer, lighter contours correspond to the $2\sigma$ (95\% confidence area) plausible areas, denoting the extended spread of the parameter space. It can be followed that the combined \emph{Hubble + BAO} dataset submits more rigorous and well disciplined contours linked to the individual datasets, efficiently reducing the parameter declinations and enhancing the general accuracy of the model constraints. This shows the mutual nature of Hubble and BAO observations in compelling cosmological parameters within the adopted theoretical framework.

\setlength{\extrarowheight}{2.5 pt}
\begin{table}[H]
\centering

\begin{tabular}{|l|l|l|}
\hline
\textbf{Data used} & \textbf{Parameters} & \textbf{Best fit values} \\ \hline

\multirow{4}{*}{Hubble} 
 & $lns$ & $3.32^{+0.21}_{-0.21}$ \\ \cline{2-3}
 & $log b_1$ & $-4.05^{+1.28}_{-1.28}$ \\ \cline{2-3}
 & $log \alpha$ & $1.83^{+2.19}_{-2.19}$ \\ \cline{2-3}
 & $\mu$ & $-0.95^{+1.02}_{-1.02}$ \\ \hline

\multirow{4}{*}{BAO} 
 & $lns$ & $9.02^{+0.17}_{-0.17}$ \\ \cline{2-3}
 & $log b_1$ & $0.70^{+2.52}_{-2.52}$ \\ \cline{2-3}
 & $log \alpha$ & $-0.01^{+2.34}_{-2.33}$ \\ \cline{2-3}
 & $\mu$ & $-0.10^{+1.63}_{-1.62}$ \\ \hline

\multirow{4}{*}{Pantheon} 
 & $c_1$ & $-1.796^{+1.274}_{-1.277}$ \\ \cline{2-3}
 & $\mu$ & $-0.741^{+1.237}_{-1.240}$ \\ \cline{2-3}
 & $log \alpha$ & $1.966^{+3.859}_{-3.865}$ \\ \cline{2-3}
 & $n$ & $1.923^{+1.019}_{-1.020}$ \\ \hline

\multirow{4}{*}{Hubble+BAO} 
 & $lns$ & $8.46^{+0.113}_{-0.113}$ \\ \cline{2-3}
 & $log b_1$ & $0.123^{+2.72}_{-3.78}$ \\ \cline{2-3}
 & $log \alpha$ & $-0.134^{+2.33}_{-2.54}$ \\ \cline{2-3}
 & $\mu$ & $0.817^{+2.01}_{-1.91}$ \\ \hline

\end{tabular}
\caption{Model fits to cosmological data, showing dataset, model name, parameter sets, and best-fit values.}
\label{tab:fit_stats}
\end{table}

\begin{table}[H]
\centering

\begin{tabular}{|c|c|c|c|}
\hline
\textbf{Data used}  & \textbf{Reduced} $\chi^2$ & \textbf{AIC} & \textbf{BIC} \\ 
\hline
 Hubble& 0.3074  &268.065  & 277.873 \\ 
\hline
 BAO& 0.3114 & 339.346   &345.179  \\ 
\hline
 Pantheon& 0.980  &3078.30  & 3112.98 \\ 
\hline
Hubble+BAO & 0.095  &879.638  &892.589  \\ 
\hline
\end{tabular}
\caption{Statistical results for different datasets.}
\label{tab:fit_stats}
\end{table}

\section{Om(z) Diagnostics}
		\label{sec4}

The $Om(z)$ parameter reads \cite{sahni2008two,Blake_2012}
\begin{equation}
    Om(z)=\frac{[\frac{H(z)}{H_0}]^2-1}{(1+z)^3-1}= \frac{\left[\frac{-b_1c_1+\frac{3b_1(log2) D(n)}{2m\alpha(n+2)(2\mu-1)}}{-b_1c_1+\frac{3b_1log(1+(1+z)^{-m}) D(n)}{2m\alpha(n+2)(2\mu-1)}}\right]^2-1}{(1+z)^3-1}
\end{equation}

where, 
$$D(n)=\left( -b_{1} \pm \sqrt{b_{1}^{2} - 4(2\mu - 1)(n - 1)} \right)
    (n - 1)^{\frac{1}{2}}$$

\begin{figure}[H]
    \centering
    \includegraphics[width=0.8\textwidth]{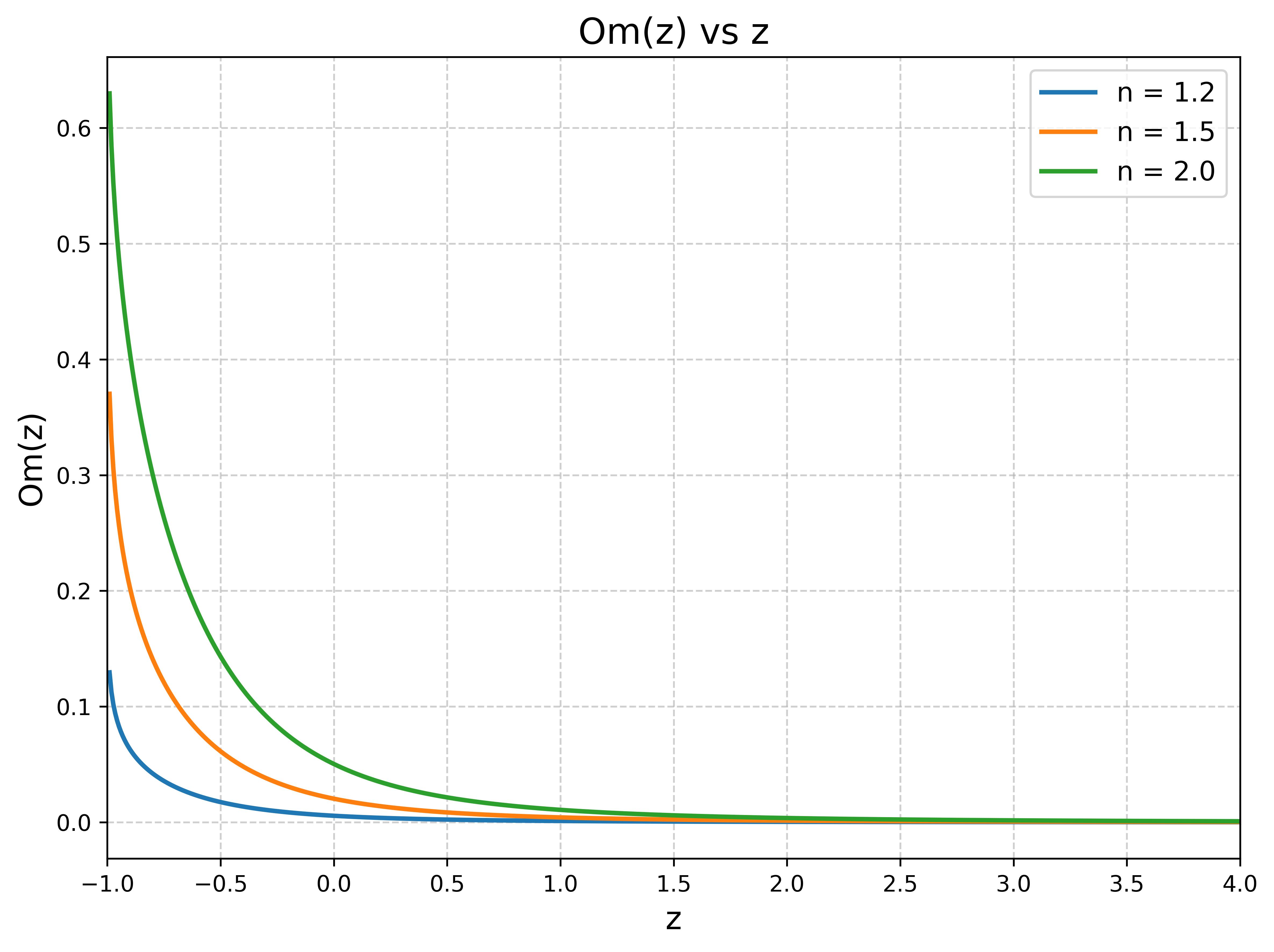}
    \caption{Evolution of the  $Om(z)$ with respect to $z$ for different values of the anisotropy parameter n.}
     \label{10}
\end{figure}

Figure~\ref{10} indicates the evolution of the parameter $Om(z)$ for three different values of the model parameter $n$ using the exact expression of $H(z)/H_0$. It is known that, if the curvature
of $Om(z)$ is positive with respect to $z$, the model is a phantom dark energy model,  for negative it is a quintessence dark energy model. For zero curvature, it represents the $\Lambda$CDM model. It shows here an increase from positive values  to larger positive values as the redshift varies from higher to lower, depicting like $\Lambda$CDM-like scenarios. Larger values of $n$ produce higher values of $Om(z)$ across the evolution, with the curve for $n=2.0$ lying highest, followed by $n=1.5$ and $n=1.2$. The difference between the curves is most commanding at low redshift ($z \approx 0$), whereas at higher redshift ($z \gtrsim 2$) the curves gradually converge as $Om(z)$ paths very small values. In general, the graph shows that increasing $n$ improves the magnitude of $Om(z)$ and modifies the action of the $Om(z)$ parameter. It can be  observed that the decreasing behavior of $Om(z)$ parameter as $z \Rightarrow 0$ indicates the quintessence like behavior of the universe.

\section{Conclusion}
		\label{sec5}

We have investigated the LRS(Locally Rationally Symmetric) Bianchi type-I universe model in $f(R,T^\psi)$ gravity. In this, we have given the model in the form of Hubble as a function of z and Potential $V$ as a function of $\psi$ (Scalar field). Paper shows the particular values for the parameters $logs$, $log b_1$, $log \alpha$, $\mu$, $c_1$,$n$ and shown graphical illustration of various parameters based on the particular values. The study which is given in this paper is summarized as an investigation of particular model, containing actions of its parameter under the specific values of arbitrary constants. Free parameter of the studied model are fitted we Observational Hubble Data (OHD), baryon acoustic oscillation (BAO), Pantheon and also combined of Hubble and BAO datasets using statistical mechanism based on MCMC method.

We have used Observational Hubble Data (OHD), baryon acoustic oscillation (BAO) and Pantheon compilation. Two dimensional confidence levels contour plots for the parameters of the deduced model are shown in Figure~\ref{3}, Figure~\ref{4}, Figure~\ref{5} and Figure~\ref{6}. The best fit values of parameters for the deduced model are tabulated in Table(1) for the diverse observational datasets. In Table(2) we shown values of reduced $\chi^2$, AIC and BIC which shows how model is statistically conservative. 

In Figure~\ref{2} we have describe the behavior of potential function $V(\psi)$ with respect to cosmic time $t$ by taking diverse values of parameter with a defined range. It concludes that the potential becomes smooth at later cosmic times because of $t$ increases, it progressively coming to zero. In Figure~\ref{3}, Figure~\ref{4}, Figure~\ref{5} and Figure~\ref{6}, we get the confidence level contours with best fit values for parameters with reduced $\chi^2$, showing that the model and observational data have great resonance. Also, derived statistical values describes the stability and consistency of the model, verifying its capacity to depict the late time acceleration precisely.

Figure~\ref{7} and Figure~\ref{8} shows the error plot for $H(z)$ model using Hubble and Pantheon datasets respectively. Particularly Figure~\ref{7} compares the $\Lambda$ CDM model cosmology with rebuilt Hubble $H(z)$ model. This indicating model's harmony with current cosmological observations by displaying reliable and accurate chronology of cosmic expansion history.

\section*{Data Availability statement}
 This research did not yield any new data.

\section*{Conflict of Interest}
Authors declare there is no conflict of interest.

\section*{Funding}
This work was supported by the Deanship of Scientific Research, Vice Presidency for Graduate Studies and Scientific Research, King Faisal University, Saudi Arabia (Funding No: KFU254133).

\section*{Acknowledgments}
MT gratefully acknowledges for JRF from  
\textbf{DST-INSPIRE Fellowship\\(IF230556), Department of Science and Technology, Ministry of Science and Technology, government of India}.~PKD would like to thank the Isaac Newton Institute for Mathematical Sciences, Cambridge, for support and hospitality during the programme Statistical mechanics, integrability and dispersive hydrodynamics where work on this paper was undertaken. This work was supported by EPSRC grant no EP/K032208/1. Also, PKD wishes to acknowledge that part of the numerical computation of this work was carried out on the computing cluster Pegasus of IUCAA, Pune, India and PKD gratefully acknowledges Inter-University Centre for Astronomy and Astrophysics (IUCAA), Pune, India for providing them a Visiting Associateship under which a part of this work was carried out. 

\bibliographystyle{elsarticle-harv}   
\bibliography{references}   

\end{document}